\documentstyle[11pt,epsf]{article}

\begin{document}

\title{\bf 
Exploring structural mechanisms in disordered materials using
the activation-relaxation technique }

\author{
G. T. Barkema$^{(a)}$,
Theoretical Physics, Utrecht University,\\
Princetonplein 5, 3584 CC Utrecht, the Netherlands. \\
Normand Mousseau$^{(b)}$, Department of Physics and Astronomy,\\
Ohio University, Athens, OH 45701, USA }

\maketitle

\begin{abstract}
Structural mechanisms in disordered materials like amorphous
semi-conductors and glasses can be explored with the
activation-relaxation technique (ART). The application of a sequence of
such mechanisms allows for the generation of well-relaxed structures.
The method and its application in the study of the microscopic changes
in amorphous silicon and silica glass are reviewed, and two recent
improvements are presented.
\end{abstract}
%\pacs{PACS: 82.20.Wt, 02.70.Rw, 61.43.Dq and 73.61.Jc.}

\section{Introduction}

While the smallest relevant time scale in atomic systems is that
belonging to atomistic oscillations, around a tenth of a pico-second,
the microscopic dynamics of glassy and complex materials at low
temperature can proceed at time scales that are easily ten orders of
magnitude larger, seconds or hours. Approaches based on the atomic
oscillation time scales, such as molecular dynamics, will not be
able to bridge this gap in the foreseeable future: alternative
approaches have to be explored.

The nature of the discrepancy between these time scales is best
understood from the configurational energy landscape: the system finds
itself in a deep minimum surrounded by energy barriers which are many
times larger than its temperature.  Only rare fluctuations of thermal
energies allow the system to jump over a barrier and move to a new
minimum.  

In this paper, we review the activation-relaxation technique
\cite{barkema96,mousseau98a} which proposes one avenue for the
exploration of such systems.  We first discuss the algorithm in some
detail, including some new algorithmic improvements, and then briefly
present its application to amorphous silicon and vitreous silica.

\section{The activation-relaxation technique}

The activation-relaxation technique consists of two parts: a path from
a local energy minimum to a nearby saddle point --- the activation; and
a trajectory from this new point to a new minimum --- the relaxation.

The relaxation to a local energy minimum is a well-defined and
well-behaved operation for which a number of efficient algorithms are
available (see, for example, Ref. \cite{numrec}).  We use an adaptive
algorithm that uses a steepest descent close to the saddle point and a
conjugate-gradient algorithm as the configuration converges to the new
minimum.

The activation from a minimum to a saddle point poses a bigger
challenge. While previous work concentrated on low-dimensional problems
for which, often, the minima were known, ART is aimed at finding saddle
points in high-dimensional energy landscapes, knowing the location of
one minimum only.

At the saddle point, all eigenvalues of the Hessian but one are
positive. In the energy landscape, this negative eigenvalue sets
the direction of a valley going down on both sides. Starting somewhere
in this valley, convergence to the saddle point can be obtained by
keeping the configuration at the energy minimum along all directions
but the one corresponding to the lowest eigenvalue, which is
assimilated with the local bottom of the valley, and pushing upward
along that one direction.  This is in essence Cerjan and Miller's
approach for locating transition states in low-dimensional energy
surfaces~\cite{cerjan81}.

This approach is too computer intensive for realistic bulk systems with
hundreds to thousands of atoms, since the time required for the
diagonalization of the Hessian grows as ${\cal O} (N^3)$.
By approximating the Hessian, ART proposes an ${\cal O} (N)$ algorithm. 
Its standard implementation follows a modified force vector
$\vec{G}$, obtained by {\it inverting} the component of the force
{\it parallel} to the displacement from the current position to the
local minimum $\vec{r}=\vec{x}-\vec{x}_m$
while minimizing all other $3N-1$ directions:
\begin{equation}
\vec{G} = \vec{F} - (1+\alpha) (\vec{F}\cdot\hat{r}) \hat{r}
\label{eq:art}
\end{equation}
where $\hat{r}$ is the normalized vector parallel to $\vec{r}$,
$\vec{F}$ is the total force on the configuration as calculated using
an interaction potential, and $\alpha$ is a control parameter.
Iteratively, this redefined force is followed until
$\vec{F}\cdot\hat{r}$ changes sign.

Because of this projection, the standard algorithm fails for valleys
perpendicular to $\vec{r}$.  We now introduce a trailing position
$\vec{x}_t$, that initially is located in the old minimum, but is moved
in the direction of the position as soon as it is more than a trailing
distance $r_t$ behind.  The direction in which we invert the force is
now chosen according to $\vec{r}=\vec{x}-\vec{x}_t$, thus avoiding the
limitation mentioned above.  This improvement is particularly helpful
in the simulation of systems with strong short-range potentials, like
metallic glasses.

Since the force $\vec{G}$ as redefined in Eq.~(\ref{eq:art}) is not
curl-free, it cannot be written as the gradient of a redefined energy
function.  For the convergence to the saddle point we therefore have to
modify the standard conjugate gradient method as for instance in
\cite{numrec}: the line minimization in direction $\hat{h}$ is replaced
by a root-finding algorithm of $\vec{G} \cdot \hat{h}$.

In non-degenerate disordered materials, only two valleys start at the
minimum: the ones corresponding to the softest vibration mode around
the minimum. To explore other valleys leading to other saddle points,
we must therefore leave the harmonic well before starting the search
for valleys. We discuss separately the direction of the initial
displacement and its size.

Any random escape direction overlaps with the softest elastic modes.
While following the redefined force, they tend to dominate exponentially
rapidly.  The softest modes should therefore be eliminated from the
initial displacement. Starting with an initially random direction
$\vec{r}_0$, we suppress the softest modes by $n$ iterations of
$\vec{r}_{i+1}=-\vec{F}(\vec{r}_i)$, a series of very small steps,
mathematically equivalent to multiple applications of the Hessian.

This suppression of the softest modes in the initial displacement has
the side-effect that it boosts the strongest modes exponentially. The
result is that the initial displacement points to a very stiff
direction, and the energy raises rapidly along this direction.  We propose
here a modification:  the stiffest directions are removed from this
vector $\vec{r}_n$ by simply making it orthogonal to directions
$\vec{r}_{n+1}, \dots, \vec{r}_{n+m}$, with $n$ and $m$ depending on
the details of the system studied.  This leaves us following a random
vector from which the softest and stiffest modes have been removed.

This initial direction is then followed until the ratio of the 
perpendicular to the parallel component of the force, as projected on 
the displacement from the minimum, reaches some ratio, 
indicating that the harmonic region has been left. 

\section{Applications}

The activation-relaxation technique is well suited to identify
atomistic mechanisms for diffusion and relaxation in disordered
systems, since the events that are created follow closely the physical
activation paths. The method was recently used for the identification
of microscopic relaxation and diffusion mechanisms in two materials:
amorphous silicon and vitreous silica.

\subsection{Amorphous silicon}

\begin{figure}
\epsfxsize=8cm
\epsfbox{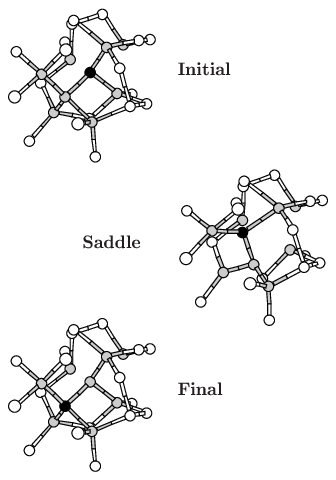}
\caption{A perfect event in amorphous silicon, corresponding to
the concerted-exchange mechanisms proposed by Pandey for 
self-diffusion in crystalline silicon. From top to bottom:
the initial, saddle point, and final configurations. The dark
atoms participate directly into the event while the white ones
are their nearest neighbors. The activation barrier is 5.12 eV,
with no asymmetry since the initial and final configuration have
exactly the same topology.}
\label{fig:si}
\end{figure}

In a simulation study of amorphous silicon, we generated more than 
8000 events from three independent runs on 1000-atom samples 
\cite{barkema98} using the empirical Stillinger-Weber potential
\cite{sw} with an increased three-body term.  The activation 
barriers for the events range from 0 to 15 eV, peaking at 4 eV. The 
number of atoms that are displaced significantly in these events (0.1 
\AA~or more) lies typically around 40, but there are usually only a 
few bonds broken or created.  A typical event, showing only atoms 
rearranging their topology and their near-neighbors, is shown in 
Fig.\ref{fig:si}.

Since in this material the list of neighbors is well-defined, we can
identify three classes of events: in {\it perfect events} four-fold
atoms exchange bonds but keep their total coordination; in {\it
conserved events}, coordination defects diffuse around, while the
overall coordination is preserved; {\it annealing events}, finally,
involve the creation or the annihilation of defects.

For the class of perfect events,  we label the atoms that change their
bonding, and construct a loop consisting of all created and broken
bonds, visited in the loop in alternating order. The sequence of the
atoms visited by this loop gives a classification of the topological
change in the bonded network.  To avoid having many labels for the same
topological reordering, all possible loops are generated, and the
alphabetically lowest classification is chosen.  Three types of perfect
rearrangements dominate the dynamics.  They correspond in order of
likelihood to (a) an exchange of neighbors between two nearby atoms,
corresponding to the Wooten-Winer-Weaire bond-exchange mechanism
introduced initially as an artificial move, (b) the exchange of two atoms,
similar to the concerted exchange mechanism introduced by
Pandey~\cite{pandey86} and illustrated in Figure \ref{fig:si}, 
and (c) a step in between where two nearest neighbors are exchanged, giving
a ``double'' Wooten-Winer-Weaire mechanism, with a shared backbone. 
We are still working on the classification of also the
conserved and annealing events.

\subsection{Vitreous silica}

\begin{figure}
\epsfxsize=8cm
\epsfbox{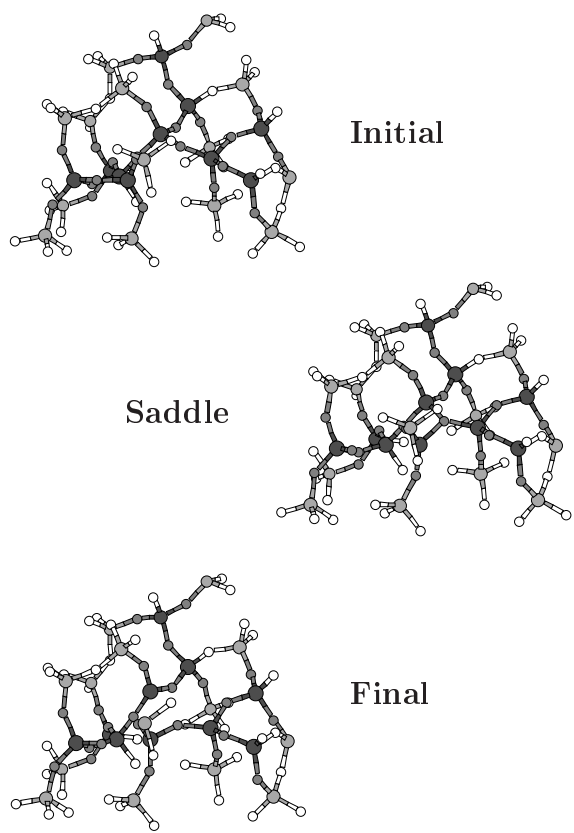}
\caption{A small event in SiO$_2$. Shown here are the  atoms
that move more than 0.1 \AA~during activation and relaxation,
plus their nearest neighbors.  Large circles are Si atoms, small ones O.
This particular event is the creation of a dangling bond on an O.
The activation barrier and asymmetry are 3.64 and 1.35 eV, respectively.
\label{fig:danglingdiff}}
\end{figure}

In a study of the mechanisms occurring in silica glass, we generated a
database of 5645 events in well-relaxed 1200-atom samples of silica, in
which the interactions were described by the screened-Coulomb potential
of Nakano {\it et al.}~\cite{nakano94}.  This study has
revealed a completely different dynamics for this material
\cite{mousseau98b} than in amorphous silicon.  In particular, because of
the need to maintain chemical ordering, the perfect mechanisms of {\it
a}-Si do not have a direct counterpart. One frequently observed
mechanism is depicted in Figure \ref{fig:danglingdiff}.

\section{Conclusion}

The activation relaxation technique provides a unique tool for
identifying the microscopic mechanisms responsible for relaxation and
diffusion in disordered materials. It avoids imposing pre-defined
atomic moves, by working in the configurational energy landscape, and
allows for real-space rearrangements of any size.  With ART, we have
been able to provide the first analysis of relaxation and diffusion
mechanisms taking place below melting in {\it a}-Si and
{\it g}-SiO$_2$ and show that their respective dynamics is 
qualitatively different.  The activation relaxation technique promises
to be a powerful tool in the investigation of a wide range complex
materials.

\section{Acknowledgements}

Part of the work reviewed here was performed in
collaboration with Simon W. de Leeuw. NM wishes to acknowledge partial
support by the NSF under Grant No. DMR 9805848.

\bibliographystyle{prsty}

%\newpage

\end{document}